\documentclass[conference]{IEEEtran}
\IEEEoverridecommandlockouts

\usepackage{cite}
\usepackage{amsmath,amssymb,amsfonts}
\usepackage{graphicx}
\usepackage{textcomp}
\usepackage{xcolor}
\usepackage{etoolbox}
\usepackage{booktabs}
\usepackage{array}%
\usepackage{listings}
\usepackage{caption}
\usepackage{tikz}
\usetikzlibrary{positioning,arrows.meta}
\usepackage{pgfplots}
\pgfplotsset{compat=1.16}
\usepackage{stfloats}%
\usepackage{float}%

\usepackage{url}
\usepackage{balance}
\usepackage[hidelinks]{hyperref}
\hypersetup{
  pdftitle={How Far Can LLMs Go in Generating Android Lint Checks from Natural Language?},
  pdfauthor={Moshood A. Fakorede, A.B. Siddique, Manu Sridharan, Umar Farooq}
}

\def\BibTeX{{\rm B\kern-.05em{\sc i\kern-.025em b}\kern-.08em
    T\kern-.1667em\lower.7ex\hbox{E}\kern-.125emX}}

\newcommand{\revision}[1]{\begingroup#1\endgroup}
\newcommand{\result}[1]{{#1}}
\newcommand{\gain}[1]{\,{\scriptsize\textcolor{green!55!black}{$\uparrow$#1}}}
\newcommand{\drop}[1]{\,{\scriptsize\textcolor{red!70!black}{$\downarrow$#1}}}
\newcommand{\pgain}[1]{\,{\scriptsize\textcolor{blue!60!black}{$\Uparrow$#1}}}
\newcommand{\bench}{\textsc{LintBench}\xspace}

\usepackage{xspace}
\newcommand{\stitle}[1]{\noindent\textbf{#1}}
\newcounter{findingctr}
\newcommand{\finding}[1]{%
  \par\vspace{3pt}\noindent\refstepcounter{findingctr}%
  \fcolorbox{black!25}{black!7}{%
    \parbox{\dimexpr\columnwidth-2\fboxsep-2\fboxrule\relax}{%
      \normalsize\itshape\textbf{Finding \thefindingctr:}~#1}}%
  \par\vspace{4pt}}
\newcommand{\myNum}[1]{(\emph{#1})}

\definecolor{fwkcolor}{rgb}{0.0,0.50,0.0}
\definecolor{platcolor}{rgb}{0.0,0.45,0.0}

\lstdefinestyle{kotlin}{
  language=Java,
  basicstyle=\ttfamily\scriptsize,
  keywordstyle=\color{blue}\bfseries,
  commentstyle=\color{gray}\itshape,
  stringstyle=\color{teal},
  breaklines=true,
  breakatwhitespace=true,
  columns=fullflexible,
  keepspaces=true,
  showstringspaces=false,
  frame=single,
  numbers=left,
  numberstyle=\tiny\color{gray},
  captionpos=b,
  xleftmargin=2.5em,
  xrightmargin=0.5em,
  framexleftmargin=2em,
  numbersep=8pt,
}
\begin{document}

\title{How Far Can LLMs Go in Generating Android Lint Checks from Natural Language?}

\author{
\IEEEauthorblockN{\href{https://orcid.org/0009-0003-2057-9865}{Moshood A. Fakorede}\IEEEauthorrefmark{1},
\href{https://orcid.org/0000-0002-3587-7289}{A.B. Siddique}\IEEEauthorrefmark{2},
\href{https://orcid.org/0000-0001-7993-302X}{Manu Sridharan}\IEEEauthorrefmark{3},
\href{https://orcid.org/0000-0001-7229-9847}{Umar Farooq}\IEEEauthorrefmark{1}}
\IEEEauthorblockA{\IEEEauthorrefmark{1}Louisiana State University,
\IEEEauthorrefmark{2}University of Kentucky\\
\IEEEauthorrefmark{3}University of California, Riverside\\
mfakor1@lsu.edu, siddique@cs.uky.edu, manu@cs.ucr.edu, ufarooq@lsu.edu}
}

\maketitle

\begin{abstract}
Static analysis tools such as Android Lint, PMD, and CodeQL let developers add custom checks for project-specific rules. 
However, authoring a check is a demanding programming task: a rule can be stated in a few sentences, yet turning it into a working, tested check requires detailed knowledge of the analysis framework and its issue-detection logic. 
Recent work generates such checks with large language models (LLMs), but each system needs more than the natural language description a developer always has: a bug-fix patch, a CVE record, or a developer-written test suite to repair against. 
No prior work measures whether an LLM can turn a description into a working check, or what external knowledge it needs to do so.

We introduce \bench, a human-validated benchmark of \result{113} real-world Android Lint checks, each paired with a natural language description and its original test suite as a held-out oracle. 
We evaluate four LLMs, including proprietary and open-weight models. 
\revision{We compare prompting strategies, then evaluate execution-guided repair with and without retrieved API knowledge.}
Prompting alone reaches \result{30\%}--\result{65\%} (pass@5) across models. 
\revision{Execution-guided repair and retrieval reach \result{57\%}--\result{85\%} test-passing rates within five calls.}
Our results show that execution feedback raises the pass rate, while retrieved framework knowledge raises the compile rate, which reaches up to \result{99\%}. 
The framework knowledge helps the low-performing models the most, providing the knowledge they need to fix compilation errors. Of the generated checks that compile but fail their tests, \result{87\%} have false negatives, missing an issue the generated check should report. 
Based on our findings, getting the issue-detection logic right remains an open challenge. 
We release \bench and the evaluation harness to support future work on generating Android Lint checks.
\end{abstract}


\section{Introduction}
\label{sec:intro}

Static analysis tools are widely used to ensure code quality, flagging bugs and security vulnerabilities by checking code against predefined rules. 
Built-in rules do not always capture a team's coding conventions, so analyzers such as Android Lint, PMD, and CodeQL let developers add custom checks for the rules they care about~\cite{androidlint, pmd, codeql}. 
Studies of developer needs report a broad demand for such project-specific rules~\cite{tymchuk2018jit, christakis2016developers}. 
Yet, each custom check is a small program written against the analyzer's framework, and authoring a check is known to be difficult~\cite{mendonca2022custom}. 
As a result, teams encode far fewer rules than they would like. 
Given the success of large language models~(LLMs) for generating code, a natural question is whether they can generate these checks too.

Recent work pursues this goal with LLMs.
KNighter synthesizes Clang Static Analyzer checkers from historical bug-fix
patches~\cite{knighter}. QLCoder synthesizes CodeQL queries from CVE metadata~\cite{qlcoder}, which typically pairs a textual description with the weakness type (CWE), the affected versions, and references such as the fixing commit.
Neither of these works builds a check from the description alone,
the one artifact a developer is sure to have (Table~\ref{tab:related}).
Closest to a description-driven setting, AutoChecker generates PMD checkers, but it is \emph{test-driven}: it provides a natural language description \emph{and} developer-written tests to the model as input, where the model is tasked to repair the checker until those tests pass~\cite{autochecker}. 
Exposing the tests during generation is not just an added burden, it also undermines validity. 
With the tests exposed to an LLM, it can reward-hack the oracle, hard-coding or special-casing the shown cases instead of implementing the rule\cite{baker2025monitoring,denison2024subterfuge}.
Thus, we argue that the tests must be kept as a held-out oracle that the model never sees.

\begin{table}[t]
\centering
\caption{
Prior LLM check synthesizers vs.\ \bench. \bench evaluates check generation from a natural-language description alone. The tests are held-out oracle, and the repair conditions receive only test outcomes.
}
\label{tab:related}
\footnotesize
\setlength{\tabcolsep}{8pt}
\begin{tabular}{@{}lll@{}}
\toprule
Approach & Analyzer & Input \\
\midrule
KNighter~\cite{knighter}       & Clang Static Analyzer & Bug-fix patch \\
QLCoder~\cite{qlcoder}         & CodeQL       & CVE metadata \\
AutoChecker~\cite{autochecker} & PMD          & Description + tests \\
\textbf{\bench} (ours)         & Android Lint & Description \\
\bottomrule
\end{tabular}
\vspace{-2em}
\end{table}

We study how far LLMs can go in this description-only setting for Android Lint, the analyzer shipped with the Android build toolchain and run by default on nearly every Android project~\cite{androidlint}. 
Besides its hundreds of built-in checks, Android Lint runs custom checks that libraries distribute to their users. To measure the adoption of custom checks, we searched Gradle build files on GitHub (\texttt{extension:gradle} and \texttt{extension:kts}) for \texttt{lintPublish}, the configuration that packages a library's custom checks with the library.
We confirmed a \texttt{lintPublish} invocation in 471 of the 537 matching files (88\%), which span 245 distinct non-fork repositories owned by 67 organizations.
It includes teams at Uber and Slack that maintain their own checks~\cite{uberlint, slacklint,composelint}. 
\revision{Android Lint requires choosing scanner interfaces and callbacks for Java and Kotlin source, XML resources, manifests, or Gradle files, then expressing detection logic through the corresponding framework APIs.}

Generating a check this way demands two capabilities: \emph{lint framework
understanding}, i.e., which scanner to implement and which Lint APIs to call, and the \emph{issue detection~logic}, which decides when to report.
Which of these two capabilities is more critical is unknown and consequential. If checks fail
for lack of framework knowledge, retrieving that knowledge should close the gap. If they fail
on the detection logic, no amount of framework documentation would help, and the bottleneck
is the model's own reasoning. We resolve this question with a benchmark of real Lint checks
paired with their tests (as held-out oracle) and a controlled study across four LLMs,
isolating what retrieval sources and execution-guided repair each contributes.

\stitle{Problem setting.} The model is given a rule's natural language description and must
synthesize a check that compiles and passes a test suite encoding the rule's intended behavior.
\revision{We keep the tests as a held-out oracle. The model cannot access the test source or
reference implementation. The description-only task specification mirrors a realistic authoring setting
where a developer can describe a rule but has not yet written its tests. Depending on the
condition, the model also receives examples, scaffolding, retrieved documentation, or compiler
and test diagnostics. We use the held-out tests to evaluate candidates and provide diagnostic
feedback during repair.}

\stitle{Benchmark.} We build \bench, \result{113} human-validated real checks from the production Android Open Source Project (AOSP)~\cite{aosp} Lint suite, each paired with a natural language description from the check's developer documentation and its test suite as an executable oracle. 
The instances cover Java/Kotlin source, XML resources, and manifests, and 23\% of them require more than one scanner interface.
We keep only checks that pass mutation-based filtering and manual validation, so a passing check
requires non-trivial issue-detection logic and every description admits a correct implementation.
A harness scores each generated check by compiling it and running the tests, providing an objective pass/fail signal.

Using \bench, we evaluate the information sources that current check-generation approaches rely on. 
We start from
\emph{prompting} on the description alone and add two richer prompts: in-context examples of well-formed checks, as KNighter uses~\cite{knighter}, and structural scaffolding that hands the model a solution skeleton to complete. 
We then add \emph{execution-guided repair}, feeding compiler and test diagnostics back to the model as in self-debugging~\cite{chen2024selfdebugging,olausson2023selfrepair}. 
This generate--run--repair loop mirrors an agentic coding workflow, where the model acts, observes execution feedback, and revises, but in a controlled, fixed-budget, fixed-tool setting that isolates each component's contribution.
Finally, we add \emph{retrieval}, which augments the prompt with external knowledge in the style of retrieval-augmented generation~\cite{lewis2020rag, apidocrag}, drawing on two sources, the Android Lint API reference~\cite{lintApiGuide} and the Lint developer guide~\cite{lintUserGuide}. 

\stitle{Research questions.} We organize the study around four questions. The first examines what prompting alone achieves, and the rest analyze the retrieval-and-repair setting:
\begin{itemize}
  \item \textbf{RQ1:} From a natural language description alone, how far does LLM prompting reach,
        and at which stage (compilation or issue-detection logic) do its failures concentrate?
  \item \textbf{RQ2:} How far does check generation reach once execution-guided repair and retrieved framework knowledge are added, and how much does each contribute to the gain?
  \item \textbf{RQ3:} When generated checks fail, is it because of a knowledge gap (missing framework APIs) or incorrect issue-detection logic?
  \item \textbf{RQ4:} What does each component (the repair loop and each retrieval source) cost per successfully generated~check?
\end{itemize}

We study these questions by evaluating each of the conditions across four LLMs, two open-weight (Qwen3.6-Max-Preview~\cite{qwen36max} and Kimi K2.7 Code~\cite{kimik27}) and two proprietary (Claude Sonnet 4.6~\cite{claudesonnet46} and Gemini 3.5 Flash~\cite{gemini35flash}).

\stitle{Findings.} Prompting from the description alone plateaus at \result{64.6\%}, with most failures concentrated at compilation~(RQ1). 
Adding execution-guided repair and retrieval raises the best result to \result{85.0\%} at a matched five-call budget, and a repair-only control shows the repair loop accounts for most of this gain. 
The top-performing models nearly saturate compilation on their own by self-correcting APIs from compiler feedback, so API-reference retrieval adds little for them. 
It helps more for the low-performing models, which correct fewer of these errors~(RQ2).
Once a check compiles but fails tests, none of the retrieved sources improves it. 
Compiling but failing checks are unsound (under-detection): checks that fail to flag the issue. 
This gap in the issue-detection logic closes only with the model's coding capability~(RQ3).
On cost, the cheapest model is not the top-performing one. 
Retrieval is roughly cost-neutral per solved check. The expanded context raises the per-run cost, but the higher pass count offsets it~(RQ4).

\stitle{Contributions.} \revision{To the best of our knowledge, \bench is the first benchmark and controlled study of generating static analysis checks from natural-language task descriptions with hidden test source~(Table~\ref{tab:related}).}
This work makes the following contributions:

\begin{itemize}
  \item \bench, a human-validated benchmark of \result{113} real Android Lint checks, each paired with a natural language description and its original test suite, with an execution harness (Section~\ref{sec:benchmark}).
  \item A controlled study across four LLMs that isolates prompting, retrieval, and
        execution-guided repair, attributes each gain to its source (execution feedback raises
        the pass rate, retrieval the compile rate), and quantifies the cost each component adds
        (Sections~\ref{sec:setup}--\ref{sec:results}).
  \item \revision{An instance-level failure analysis separating under-detection, runtime errors,
        and message or location failures (Sections~\ref{sec:results} and~\ref{sec:discussion}).}
\end{itemize}

\section{Anatomy of a Lint Check}
\label{sec:background}

Android Lint is the standard static analyzer for Android, extensible with developer-written
checks.

\stitle{Terminology.} \revision{Four} framework terms recur throughout this paper. A \revision{\emph{check}} is a
single rule Lint enforces. It is implemented as a \revision{\texttt{Detector}}, the Java or Kotlin
class that runs the analysis, which declares one or more \revision{\texttt{Issue}}s, the metadata
(id, message, severity, category) for each problem it reports, and implements a
\revision{\emph{scanner}} interface (e.g., \texttt{SourceCodeScanner} or
\texttt{XmlScanner}) that fixes the program representation it inspects.
One detector may declare several issues, so we identify a check by its \texttt{Issue}, not its
detector. Generating a check means writing a detector that registers the right issue and implements
the right scanner.

Because the targeted representation varies, the same intent, ``flag $X$,'' yields very different
code depending on whether $X$ lives in source, a layout, the manifest, or the build script.
Consider Lint's own \texttt{SecureRandom} check, whose developer-facing
description reads: \emph{``Using a fixed seed with SecureRandom \ldots will cause the instance to
return a predictable sequence of numbers \ldots not appropriate for secure use.''}
\label{sec:motivating} The description states the rule's intent and the JDK class it concerns,
but nothing about how to express it as a Lint check (Listing~\ref{lst:securerandom};
\bench, Section~\ref{sec:descriptions}).

\begin{lstlisting}[float,language=Java,abovecaptionskip=4pt,belowcaptionskip=-20pt,
emph={Detector,SourceCodeScanner,getApplicableMethodNames,visitMethodCall,JavaContext,UCallExpression,PsiMethod,isMemberInClass,ConstantEvaluator,evaluate,evaluator,valueArguments,report,getLocation,Issue,create,Implementation,Scope,Severity,Category},
emphstyle=\bfseries\color{fwkcolor},
stringstyle=\color{black},
commentstyle=\itshape\tiny\color{gray},
caption={Condensed \texttt{SecureRandom} check (Kotlin). Everything in
\textcolor{fwkcolor}{green} is Lint framework code: the scanner interface, callbacks, and API
utilities the model must implement. The rule itself touches only an ordinary JDK target (black,
\texttt{setSeed} on \texttt{java.security.SecureRandom}).},
label={lst:securerandom}]
class SecureRandomDetector :
    Detector(), SourceCodeScanner {
  override fun getApplicableMethodNames() =
    listOf("setSeed")
  override fun visitMethodCall(context: JavaContext,
      node: UCallExpression, method: PsiMethod) {
    val args = node.valueArguments
    if (args.isEmpty()) return
    if (!context.evaluator.isMemberInClass(method,
          "java.security.SecureRandom")) return
    if (ConstantEvaluator.evaluate(context, args[0]) != null)
      context.report(ISSUE, node,
        context.getLocation(node), MSG)
  }
  // companion: ISSUE = Issue.create("SecureRandom", ...)
}
\end{lstlisting}

\stitle{What a working check demands.}\label{sec:why-hard} \looseness=-1 The rule takes only a sentence or two to
state. A working check must still supply four elements the description leaves
implicit. \myNum{i}~\emph{Choosing the scanner and scope}: which representation to scan and which
callbacks to implement, here a \texttt{SourceCodeScanner} intercepting the \texttt{setSeed} call.
\myNum{ii}~\emph{Grounding the framework API}: precise utilities such as
\texttt{JavaEvaluator.isMemberInClass} and \texttt{ConstantEvaluator}, called with their exact
signatures. \myNum{iii}~\emph{Assembling and registering the check}: declaring an \texttt{Issue}
and wiring its \texttt{Implementation} and registry entry by framework convention.
\myNum{iv}~\emph{Getting the issue-detection logic right}: confirming the receiver is a
\texttt{SecureRandom} and deciding whether the seed is a compile-time constant. Our study measures
how far current LLMs reach on these demands, and where they fall short
(Section~\ref{sec:results}).

\begin{figure}[!t]
\centering
\includegraphics[width=\columnwidth]{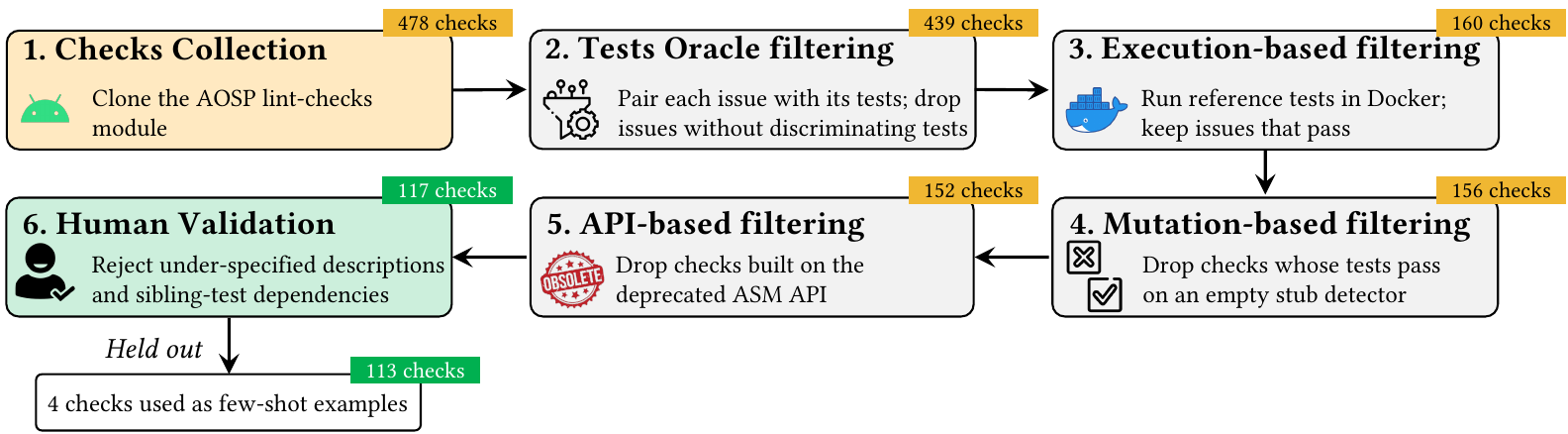}
\caption{\bench construction. Four filters reduce \result{478} candidate issues to \result{152};
human validation flags \result{35} and \result{four} are held out as few-shot examples, yielding the
\result{113}-check \bench. Badges show the count after each step.}
\label{fig:construction}
\vspace{-1.5em}
\end{figure}

\section{\bench}
\label{sec:benchmark}

\bench is, to the best of our knowledge, the first public benchmark for generating Android Lint
checks from a natural-language description. Its \result{113} instances are human-validated real
checks from the production AOSP detector suite, each paired with the developer's description and the
check's own test suite as an executable oracle. Together, the instances cover source, XML resource, and manifest checks. Three properties make it a realistic and discriminating benchmark.
\myNum{i}~\emph{Realistic:} every check ships in Android's official tooling.
\myNum{ii}~\emph{A behavior-validated oracle:} we keep an instance only after confirming that its tests fail on an empty stub detector.
\myNum{iii}~\emph{Human-validated:} every description is reviewed for clear specification. 

\subsection{Dataset Construction}

We construct \bench in the steps of Figure~\ref{fig:construction}.

\stitle{1.~Checks collection.} We clone the AOSP \texttt{lint-checks} module~\cite{aosp}, the canonical
set of production Android Lint detectors, and pair each lint issue with the JUnit tests that exercise
it (a detector may define several issues), yielding \result{478} candidate issues whose own tests
serve as an executable correctness oracle.

\stitle{2.~Tests-oracle filtering.} We keep the \result{439} issues backed by a discriminating oracle:
\emph{per detector}, at least two test methods, ideally checking both flagged and clean code
(dropping \result{32} issues with absent or too-thin tests); \emph{per issue}, the issue's own ID
asserted in some test, so the signal is specific to that check and not a sibling (dropping
\result{7} more).

\stitle{3.~Execution-based filtering.} We run each issue's tests against its reference implementation in
an isolated, version-pinned Docker harness, keeping the \result{160} that pass; the rest depend on
AOSP-internal utilities, shared fixtures, or cross-module helpers unavailable in the harness.

\stitle{4.~Mutation-based filtering.} We replace each reference detector with an empty stub and rerun
its tests, keeping only checks whose tests catch this mutant. \result{Four} still pass and cannot
tell a working detector from one that does nothing, leaving \result{156}.

\stitle{5.~API-based filtering.} We drop \result{four} checks built on the deprecated ASM
bytecode-visitor API, restricting \bench to the current UAST source-analysis API (the unified syntax
tree Lint exposes for Java and Kotlin)~\cite{uast}, leaving \result{152} valid~checks.

\begin{table}[t]
\centering
\footnotesize
\setlength{\tabcolsep}{3pt}
\caption{Composition of \bench and the unverified \result{148} set. Each group
partitions the benchmark independently.}
\vspace{-0.5em}
\label{tab:diversity}
\begin{tabular}{@{}lrr@{}}
\toprule
 & \bench-unverified & \bench \\
 & (148) & (113) \\
\midrule
\multicolumn{3}{@{}l}{\textit{Analysis representation}} \\
\quad XML / manifest (\texttt{XmlScanner})               & 67 (45\%) & 54 (48\%) \\
\quad Source (\texttt{SourceCodeScanner})                & 41 (28\%) & 33 (29\%) \\
\quad Mixed (source $+$ XML)                             & 38 (26\%) & 26 (23\%) \\
\quad Other (Gradle, binary, \ldots)                     & 2 (1\%)   & 0 (0\%) \\
\midrule
\multicolumn{3}{@{}l}{\textit{Implementation language}} \\
\quad Java                                               & 91 (61\%) & 63 (56\%) \\
\quad Kotlin                                             & 57 (39\%) & 50 (44\%) \\
\midrule
\multicolumn{3}{@{}l}{\textit{Difficulty}} \\
\quad Easy                                               & 41 (28\%) & 38 (34\%) \\
\quad Medium                                             & 82 (55\%) & 61 (54\%) \\
\quad Hard                                               & 25 (17\%) & 14 (12\%) \\
\midrule
\multicolumn{3}{@{}l}{\textit{Issue category}} \\
\quad Correctness                                        & 87 (59\%) & 71 (63\%) \\
\quad Performance                                        & 18 (12\%) & 11 (10\%) \\
\quad Icons                                              & 15 (10\%) & 8 (7\%) \\
\quad Security                                           & 9 (6\%)   & 9 (8\%) \\
\quad Other (8 categories)                               & 19 (13\%) & 14 (12\%) \\
\bottomrule
\end{tabular}
\vspace{-2em}
\end{table}

\stitle{6.~Human validation.} \revision{
We manually reviewed the description and targeted tests of all 152 checks against two criteria and rejected 35, which leaves 117.
We flagged 22 as \emph{under-specified}: the description
omitted behavior or contracts required by the tests. For example, \texttt{InvalidAnalyticsName}
refers to external Firebase documentation without stating the tested naming rules.
We flagged 24 for \emph{sibling-test burden}, where passing requires implementing other
issues in the same detector. Eleven checks met both criteria. For example,
\texttt{IconDuplicatesConfig}'s \texttt{testNoDpi} also requires reports for
\texttt{IconNoDpi} and \texttt{IconMissingDensityFolder}.}

\stitle{Held-out examples.} We reserve \result{four} of the \result{117} as few-shot examples,
leaving \result{113} as \bench. \bench plus the \result{35} filtered checks in human validation form the \result{148}-check
\emph{unverified} set that Table~\ref{tab:diversity} contrasts with \bench.
\revision{The four examples, 113 benchmark checks, and 35 rejected checks are mutually disjoint.}

\subsection{Instance Structure and Task Descriptions}
\label{sec:descriptions}

Each \bench instance pairs a natural-language specification with the scanner
interface(s) the check must implement and links to the relevant Lint API reference.
The specification is derived from the check's developer-facing \texttt{briefDescription}
and \texttt{explanation} in the AOSP source. It conveys the rule's intent and, like that
documentation, may name the type or API the rule concerns (for example,
\texttt{SecureRandom}), but it never includes the check's implementation: the scanner to
extend, the callbacks to override, or the \texttt{Issue} registration. This keeps the
task (``what to detect'') separate from the solution (``how to implement it as a Lint
check''), so the base condition measures generation from intent, not from a
leaked implementation.
Full schema and field definitions are included in the released artifact.

\subsection{Dataset Characteristics}

\bench comprises \result{113} checks with \result{296} targeted test methods (mean \result{2.6} per
check); their \result{81} reference detectors span \result{50}--\result{1{,}732} source lines of code (SLoC) (median
\result{120}) and \result{11} issue categories, dominated by \textsc{Correctness} (\result{62.8\%}).
Table~\ref{tab:diversity} contrasts \bench with the larger \result{148}-check unverified set, which
tracks it closely. The mix is heterogeneous: \result{26} checks (\result{23\%}) implement more than
one scanner interface, most often source and XML together, forcing a model to coordinate two scanners
at once.

\stitle{Difficulty.} \revision{For reference detector $d$, let $x_1(d)$ be SLoC,
$x_2(d)$ its scanner-interface count, and $x_3(d)$ its test-method count across distinct test
files. We compute per-detector medians $m_j$ on the 148-check unverified set:
$m_1=123$ logical source lines, $m_2=1$ scanner interface, and $m_3=6$ test methods
per detector, counting distinct test files. We retain these medians after human filtering.
The equally weighted score is
\[
q(d)=\frac{1}{3}\sum_{j=1}^{3}\mathbf{1}[x_j(d)>m_j].
\]
All checks from a detector inherit its score. The tiers are \textsc{easy} ($q=0$),
\textsc{medium} ($q\in\{1/3,2/3\}$), and \textsc{hard} ($q=1$), yielding the reported
0.33/0.67 cutpoints directly from the three indicators. We fixed these tiers before
the repair experiments.
Test count is a proxy for tested behavioral breadth. The tiers describe reference-code
complexity, and we evaluate their association with generation outcomes in
Section~\ref{sec:difficulty}.}

\subsection{Evaluation Harness}
\label{sec:harness}

The harness takes a generated check, compiles it against the Lint API, and runs the original tests
that assert its issue, recording: \myNum{i}~compile success;
\myNum{ii}~test outcome (all targeted tests pass / some fail); \myNum{iii}~compiler
and test logs for failure analysis. Before execution, two pre-processing steps are
applied. \textit{Stub injection} aligns the generated \texttt{Issue} field names with
the pre-compiled \texttt{BuiltinIssueRegistry}, preserving issue-detection logic while
allowing the JVM to initialize. \revision{\textit{Loose-mode patching} replaces supported
exact-output assertions with \texttt{expectContains("[IssueId]")} for the expected issue
IDs and removes \texttt{expectFixDiffs} calls}, so evaluation targets issue-detection logic, not
message wording. \revision{These assertions require a report for each
expected ID but do not enforce its location, multiplicity, wording, or the absence of
extra reports. Existing \texttt{expectClean()} assertions remain unchanged and reject
reports on the clean inputs they cover. This coverage varies across checks. We consider a
generated check \emph{test-passing} if it compiles and passes all targeted tests under
these rules.}

Producing framework-correct, compilable
code is a key part of authoring a Lint check: the check must call the scanner APIs
with the right signatures and assemble a registrable \texttt{Issue}. We therefore record compilation and test-passing as
separate outcomes (compile@$k$ and pass@$k$; Section~\ref{sec:setup}), which lets us locate
whether a generation fails at framework grounding or at issue-detection logic.

\stitle{Docker-isolated execution.} The pre-processed detector and test files are
mounted into a Docker container built from a Gradle environment containing
\texttt{lint-api}, \texttt{lint-tests}, and \texttt{android.jar} at pinned versions
(Android Gradle Plugin 31.7.0). Gradle compiles the detector and test class, then
runs the targeted test class with \texttt{-tests}. Compilation errors are extracted
from the build log; JUnit XML output is parsed to determine per-method pass/fail
status. The container enforces a 120-second timeout and 3\,GB memory limit.

\section{How Far Prompting Alone Reaches}
\label{sec:setup}\label{sec:baseline}

We measure how far current LLM prompting reaches on \bench from the description alone, establishing
the gap a method must close. This section first defines the apparatus for the baseline study
(RQ1), the models, metrics, and prompting strategies we compare, then reports the results.

\subsection{Models and metrics}

We evaluate four recent LLMs spanning a range of coding performance: Claude Sonnet
4.6~\cite{claudesonnet46}, Gemini 3.5 Flash~\cite{gemini35flash},
Qwen3.6-Max-Preview~\cite{qwen36max}, and Kimi K2.7 Code~\cite{kimik27}.
\revision{We selected a leading model from each of Anthropic, Google, Qwen, and Moonshot
using Android Bench rankings during May--June 2026~\cite{androidbench}, prioritizing provider diversity.
We also considered OpenRouter pricing~\cite{openrouter}: Opus 4.6 offered little ranking
improvement over Sonnet 4.6 at several times its price.}

We report two @$k$ metrics, each the probability that at least one of $k$ independently sampled
generations meets a criterion, using the standard unbiased estimator
$\text{metric@}k = \mathbb{E}[1 - \binom{n-c}{k}/\binom{n}{k}]$~\cite{humaneval,soleval}, with $n
\geq k$ samples per instance and $c$ meeting it. The criterion is \emph{compile} (the check
compiles) or \emph{pass} (it compiles and all targeted tests succeed), so compile@$k \geq$ pass@$k$;
compile isolates framework-grounding from logic failures. We report each at $k{=}1$ and $k{=}n{=}5$,
and \#Pass, the number of checks solved by at least one of the $k$~samples.

\begin{lstlisting}[float=t,language=Java,abovecaptionskip=4pt,belowcaptionskip=-20pt,
emph={Detector,SourceCodeScanner,Issue,create,JavaContext,UCallExpression,PsiMethod},
emphstyle=\bfseries\color{fwkcolor},
deletekeywords={for,return,in,is,it,to},
escapeinside={(@}{@)},
emph={[2]TODO},emphstyle={[2]\bfseries\color{red!75!black}},
stringstyle=\color{black},
commentstyle=\itshape\tiny\color{gray},
caption={Structural scaffolding prompt (C0\,+\,C2) for the \texttt{SecureRandom} check. Given the
spec and a metadata-derived skeleton (\textcolor{fwkcolor}{green}), the model fills the
\texttt{TODO()} stubs with issue-detection logic (\textcolor{red!75!black}{red}).},
label={lst:scaffold}]
System prompt:
You are an expert Android developer specialising in
Android Lint custom checks. You write (@Detector@)
implementations in Kotlin that are correct, idiomatic,
and compile cleanly against the Lint API.
// Rules are provided here (omitted)

User prompt:
Complete the following Android Lint (@Detector@) skeleton
in Kotlin.
(@Issue@) ID: SecureRandom   Class: SecureRandomDetector
Category: SECURITY   Severity: WARNING
Base (@class@): (@Detector@)

Specification: Using a fixed seed with `SecureRandom`
Specifying a fixed seed will cause the instance to
return a predictable sequence of numbers. This may be
useful for testing but it is not appropriate for
secure use.

Starter skeleton -- fill in all TODO() stubs,
add helpers as needed:
class SecureRandomDetector : Detector(), SourceCodeScanner {
  companion object {
    <IMPLEMENTATION; ISSUE = Issue.create("SecureRandom", ...)>
  }
  override fun getApplicableMethodNames(): List<String>? = TODO()
  override fun visitMethodCall(context: JavaContext,
      node: UCallExpression, method: PsiMethod) { TODO() }
}

Output the complete SecureRandomDetector.kt with all
methods implemented.
\end{lstlisting}

\subsection{Prompting conditions}

We compare three prompt configurations that vary the knowledge provided to the
model, each instance sampled $k{=}5$ times at temperature $T{=}0.8$:

\begin{itemize}
  \item \textbf{Zero-shot} (C0): the natural-language description only, with no
        additional context. This measures raw LLM performance on the task.
  \item \textbf{Few-shot} (C0\,+\,C1): \revision{the description plus worked
        \texttt{Detector} examples matched to the target's scanner interfaces and language.
        The reported experiments use only four held-out source/XML examples: one Java and one Kotlin example
        for each of \texttt{SourceCodeScanner} and \texttt{XmlScanner}. These are the only interfaces
        used by 111 of the 113 checks. Each example compiles, passes its original tests,
        has fewer than 300 LOC with no hidden dependencies, and represents a distinct
        implementation pattern. Multi-interface targets receive matching examples.}
  \item \textbf{Structural scaffolding} (C0\,+\,C2): the description plus scaffolding
        derived from the target check's metadata: the \texttt{Detector} base class,
        scanner interface(s), and \texttt{Issue}/\texttt{Implementation} registration
        boilerplate. The system prompt casts the model as an expert Android Lint
        developer (Listing~\ref{lst:scaffold}): it is handed the framework basics and
        asked only for the check-specific issue-detection logic, which it must still produce in full.
\end{itemize}

\subsection{Baseline results}

Zero-shot
pass@1 ranges from \result{13.3\%} (Kimi) to \result{34.3\%} (Gemini), and even at five
samples the best prompt setting solves only \result{64.6\%} of checks
(Table~\ref{tab:baseline-results}).
Implementing the rule as compilable code is the challenge: most first attempts
never compile, especially for the open-weight models (compile@1
\result{24.6\%} for Kimi and \result{26.5\%} for Qwen, against \result{55}--\result{63\%} for
the proprietary Claude and Gemini). Compile-stage errors account for roughly three-quarters of the
open-weight models' samples (\result{74\%} Qwen, \result{75\%} Kimi). The dominant compile failure is API hallucination, which we flag when a sample's compiler
diagnostic reports an unresolved symbol or an invalid override against the Lint or UAST API.
It accounts for \result{53\%} of zero-shot
compile failures (\result{31\%} of all generations) and falls monotonically with model performance,
from \result{41\%} of Kimi's generations to \result{17\%} for Claude, but never vanishes.
Structural scaffolding, which hands the model the scanner and \texttt{Issue} skeleton, lowers
hallucination for three of the four models (Qwen \result{40}$\to$\result{22\%}, Kimi
\result{41}$\to$\result{28\%}, Gemini \result{25}$\to$\result{14\%}) but not for Claude
(\result{17}$\to$\result{23\%}), indicating that this hallucination pattern drives its uneven
effect on pass@1.

\begin{table}[t]
\centering
\small
\setlength{\tabcolsep}{4pt}
\caption{Prompting baselines on \bench ($k=5$ samples). Arrows indicate the change vs.\ the zero-shot.}
\label{tab:baseline-results}
\resizebox{\columnwidth}{!}{%
\begin{tabular}{@{}lccccc@{}}
\toprule
Setting + LLM & compile@1 & pass@1 & compile@$k$ & pass@$k$ & \#Pass \\
\midrule
\multicolumn{6}{@{}l}{\textbf{Zero-shot}\hfill\textit{natural-language description only}}\\
\quad + Qwen3.6-Max-Preview & 26.5 & 14.3 & 54.9 & 31.9 & 36 \\
\quad + Kimi K2.7 Code      & 24.6 & 13.3 & 61.1 & 38.1 & 43 \\
\quad + Claude Sonnet 4.6   & 62.5 & 33.1 & 85.0 & 47.8 & 54 \\
\quad + Gemini 3.5 Flash    & 55.2 & 34.3 & \textbf{89.4} & 61.1 & 69 \\
\midrule
\multicolumn{6}{@{}l}{\textbf{Few-shot}\hfill\textit{with worked example checks}}\\
\quad + Qwen3.6-Max-Preview & 33.1 & 15.8 & 55.8\gain{0.9} & 30.1\drop{1.8} & 34 \\
\quad + Kimi K2.7 Code      & 34.2 & 18.8 & 57.5\drop{3.6} & 36.3\drop{1.8} & 41 \\
\quad + Claude Sonnet 4.6   & 54.0 & 31.3 & 67.3\drop{17.7} & 42.5\drop{5.3} & 48 \\
\quad + Gemini 3.5 Flash    & 38.1 & 24.6 & 61.9\drop{27.5} & 38.9\drop{22.2} & 44 \\
\midrule
\multicolumn{6}{@{}l}{\textbf{Structural scaffolding}\hfill\textit{scanner and issue skeleton}}\\
\quad + Qwen3.6-Max-Preview & 48.0 & 23.0 & 69.9\gain{15.0} & 37.2\gain{5.3} & 42 \\
\quad + Kimi K2.7 Code      & 42.8 & 22.1 & 74.3\gain{13.2} & 48.7\gain{10.6} & 55 \\
\quad + Claude Sonnet 4.6   & 47.8 & 26.4 & 67.3\drop{17.7} & 40.7\drop{7.1} & 46 \\
\quad + Gemini 3.5 Flash    & \textbf{66.7} & \textbf{42.3} & 84.1\drop{5.3} & \textbf{64.6}\gain{3.5} & \textbf{73} \\
\bottomrule
\end{tabular}%
}
\vspace{-2em}
\end{table}

Richer prompts move this ceiling little. Structural scaffolding raises pass@1 for three of the four models (for example, Qwen 14.3\%$\rightarrow$23.0\% and Gemini 34.3\%$\rightarrow$42.3\%) but lowers it for Claude (33.1\%$\rightarrow$26.4\%), and few-shot examples raise pass@1 only for Qwen and Kimi. The best pass@5 rises from 61.1\% (zero-shot) to 64.6\% (Gemini with scaffolding).
These swings share one cause: added context helps a model that lacks structure and constrains one
that does not. The open-weight models start far lower (zero-shot compile@1 \result{24.6\%} Kimi,
\result{26.5\%} Qwen), so the skeleton supplies the boilerplate they cannot write and raises their
compile and pass rates. The proprietary models gain little structurally and regress at the compile
stage, as the worked example pulls in API patterns from a mismatched detector; few-shot also makes
the five samples alike, lowering pass@5 even where pass@1 rises. Claude, which already writes
compilable detectors in the zero-shot setting, only loses from the added constraints.
The deeper barrier underneath these swings is API hallucination. A hallucinated symbol is a
missing piece of Android Lint framework knowledge that the compiler error pinpoints exactly, so it
can be fixed two ways: by supplying that knowledge (retrieval) or by letting the model read its own
diagnostic and self-correct (execution-guided repair). We pursue both in
Section~\ref{sec:approach}.

\finding{Prompting from the description alone plateaus at \result{64.6\%} (pass@5). Richer prompts do not break the ceiling. The barrier is compilation, dominated by API hallucination that falls with model performance but never vanishes.}

\section{Execution-Guided Repair and Retrieval}
\label{sec:approach}

\subsection{Overview}
\begin{figure}[t]
\centering
\includegraphics[width=\columnwidth]{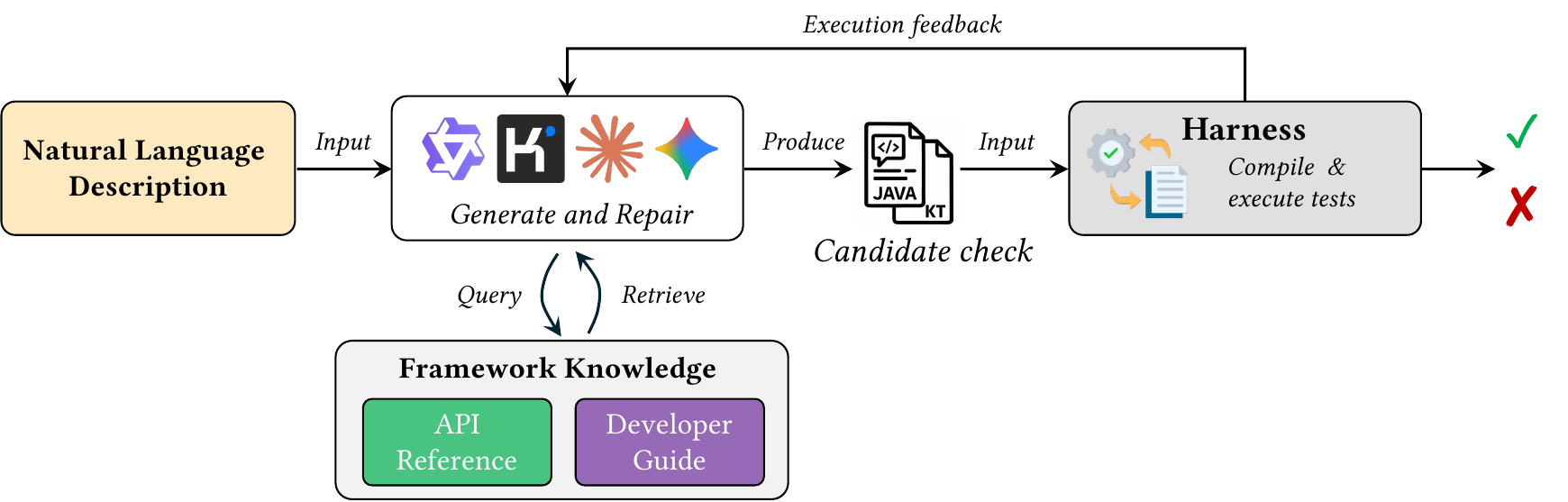}
\caption{Execution-guided repair and retrieval. The description (plus retrieved knowledge) goes to the LLM; the harness compiles and tests the candidate, feeding diagnostics back.}
\label{fig:pipeline}
\vspace{-1em}
\end{figure}

Prompting alone plateaus, with most failures never compiling (Section~\ref{sec:baseline}).
\revision{We therefore evaluate execution-guided repair, with and without retrieval of Lint
API-reference entries and developer-guide passages.} The augmented description is given to the LLM, which emits a
complete check (a \texttt{Detector} with its \texttt{Issue} registration). The harness compiles it
and runs its tests (Section~\ref{sec:benchmark}), and on failure feeds back the compiler errors and
the failing-test \emph{assertions} (the JUnit failure messages, not the test source) for repair
until the check passes or the budget is exhausted. The
model never sees the test code or a reference implementation, working only from the description and
these diagnostics (Figure~\ref{fig:pipeline}). 
This generate--run--repair loop mirrors an agentic coding workflow, in which the model acts, observes execution feedback, and revises, but it runs with a fixed call budget and fixed decoding settings, so the retrieval sources can be compared under identical repair conditions.

\subsection{Conditions and prompts}
\label{sec:conditions}
\revision{All repair conditions use the description (C0). We vary retrieval of the API reference (C3)
and developer guide (C4), including a Repair-only control with neither source. The few-shot examples
(C1) and scaffolding (C2) are evaluated separately in Section~\ref{sec:setup}.}
To prevent leakage, each target's own source, tests, and \texttt{Issue}
metadata are excluded from every pool and index.

\stitle{C3: API-reference retrieval.} We index the public Lint analysis API and the UAST/PSI
source-syntax APIs it uses ($\approx153$ entries), a scanner-interface method index ($\approx43$),
and \texttt{SdkConstants} mined from existing detectors. Each entry pairs an API element with its
signature, doc comment, and a usage snippet (target excluded), embedded with
\texttt{BAAI/bge-large-en-v1.5}~\cite{bge} and indexed with FAISS. We query with the description and
insert the top-5 entries above a 0.6 cosine threshold.

\stitle{C4: Developer-guide retrieval.} We chunk the official Lint developer guide by section (223
chunks), embed it with the same model, and insert the top-20 chunks above the same 0.6~threshold.

\stitle{Prompt construction.}\label{sec:prompts} The template includes the description and an
instruction to emit the check in a fixed output format the harness can extract. \revision{Decoding is held
fixed across the four repair conditions (Section~\ref{sec:lintgen-setup}).} We release the full template and one filled
instance per condition with the artifact.

\stitle{Scanner-conditioned, error-augmented retrieval.} A detector's structure follows the
scanner interface it implements, which the model has not chosen at the first iteration. The
initial prompt therefore opens with a fixed overview of the available scanner interfaces and
project-level callbacks and asks the model to select those the check needs. We parse the
interface(s) the generated detector actually implements and condition later retrieval on them, so
the signatures supplied follow the model's own design. Before each repair we re-retrieve with a
query augmented by the failure's compiler errors or failing-test output, biasing results toward
the symbols relevant to the fault.

\begin{table*}[t]
\centering
\footnotesize
\setlength{\tabcolsep}{7pt}
\renewcommand{\arraystretch}{1.12}
\caption{Retrieval-and-repair results on \bench. Best per column \textbf{bold}.
Arrows (matched five-call budget): Repair-only vs.\ best prompting
(\textcolor{blue!60!black}{$\Uparrow$}); retrieval vs.\ Repair-only
(\textcolor{green!55!black}{$\uparrow$}/\textcolor{red!70!black}{$\downarrow$}).}
\label{tab:lintgen-results}
\begin{tabular}{@{}lcccccc@{}}
\toprule
Setting + LLM & Compile (\%)\,$\uparrow$ & Pass (\%)\,$\uparrow$ & \#Pass\,(/113)\,$\uparrow$ & Iters\,$\downarrow$ & \$/solved\,$\downarrow$ & kTok/solved\,$\downarrow$ \\
\midrule
\multicolumn{7}{@{}l}{\textbf{Repair-only}\hfill\textnormal{\scriptsize($\Delta$ vs.\ best prompting)}}\\
\quad + Qwen3.6-Max-Preview & 74.3\pgain{4.4} & 50.4\pgain{13.2} & 57 & 3.6 & 0.308 & 55.6 \\
\quad + Kimi K2.7 Code      & 77.0\pgain{2.7} & 49.6\pgain{0.9}  & 56 & 3.8 & 0.327 & 104.3 \\
\quad + Claude Sonnet 4.6   & 93.8\pgain{8.8} & 66.4\pgain{18.6} & 75 & 2.9 & 0.152 & 16.4 \\
\quad + Gemini 3.5 Flash    & 95.6\pgain{6.2} & 80.5\pgain{15.9} & 91 & 2.5 & 0.227 & 28.1 \\
\midrule
\multicolumn{7}{@{}l}{\textbf{Repair + API reference}\hfill\textnormal{\scriptsize($\Delta$ vs.\ Repair-only)}}\\
\quad + Qwen3.6-Max-Preview & 92.0\gain{17.7} & 56.6\gain{6.2}  & 64 & 3.4 & 0.265 & 50.1 \\
\quad + Kimi K2.7 Code      & 91.2\gain{14.2} & 63.7\gain{14.1} & 72 & 3.2 & 0.202 & 64.9 \\
\quad + Claude Sonnet 4.6   & 97.3\gain{3.5}  & 73.5\gain{7.1}  & 83 & 2.6 & 0.129 & 15.2 \\
\quad + Gemini 3.5 Flash    & 98.2\gain{2.6}  & 82.3\gain{1.8}  & 93 & \textbf{2.4} & 0.219 & 28.3 \\
\midrule
\multicolumn{7}{@{}l}{\textbf{Repair + Developer guide}\hfill\textnormal{\scriptsize($\Delta$ vs.\ Repair-only)}}\\
\quad + Qwen3.6-Max-Preview & 76.1\gain{1.8} & 45.1\drop{5.3}  & 51 & 3.7 & 0.348 & 63.5 \\
\quad + Kimi K2.7 Code      & 82.3\gain{5.3} & 61.1\gain{11.5} & 69 & 3.5 & 0.228 & 71.6 \\
\quad + Claude Sonnet 4.6   & 94.7\gain{0.9} & 69.0\gain{2.6}  & 78 & 2.9 & 0.138 & 15.6 \\
\quad + Gemini 3.5 Flash    & \textbf{99.1}\gain{3.5} & 81.4\gain{0.9}  & 92 & \textbf{2.4} & 0.215 & 27.2 \\
\midrule
\multicolumn{7}{@{}l}{\textbf{Repair + API reference + Developer guide}\hfill\textnormal{\scriptsize(full; $\Delta$ vs.\ Repair-only)}}\\
\quad + Qwen3.6-Max-Preview & 94.7\gain{20.4} & 56.6\gain{6.2}  & 64 & 3.5 & 0.260 & 49.5 \\
\quad + Kimi K2.7 Code      & 92.0\gain{15.0} & 69.0\gain{19.4} & 78 & 3.3 & 0.172 & 56.0 \\
\quad + Claude Sonnet 4.6   & \textbf{99.1}\gain{5.3}  & 75.2\gain{8.8}  & 85 & 2.7 & \textbf{0.120} & \textbf{14.6} \\
\quad + Gemini 3.5 Flash    & \textbf{99.1}\gain{3.5} & \textbf{85.0}\gain{4.5} & \textbf{96} & \textbf{2.4} & 0.205 & 26.7 \\
\bottomrule
\end{tabular}
\vspace{-1.5em}
\end{table*}

\subsection{Experimental setup}
\label{sec:lintgen-setup}\label{sec:lintgen-configs}
We evaluate a Repair-only control (execution-guided repair with no retrieval) and three
retrieval configurations that pair the base description and repair with a source:
Repair + API reference (C0\,+\,C3), Repair + Developer guide (C0\,+\,C4), and
Repair + Both sources (C0\,+\,C3\,+\,C4, the full configuration). We run all four on the
same four models as the baseline study (Section~\ref{sec:setup}), each as one initial generation
plus up to four execution-guided repair attempts (five model calls total) at temperature
$T{=}0.2$, a 32{,}768-token output cap, and a 180-second per-iteration build timeout.
\revision{We chose the five-call budget after an initial ten-call pilot in which performance plateaued
at five calls. Section~\ref{sec:results} reports a further ten-call check on Gemini's remaining failures.}

\stitle{Pass rate and compile rate.} \emph{Pass rate} is the fraction of instances whose single run
yields a passing check (the pass criterion of Section~\ref{sec:setup}); \emph{compile rate}, the
fraction whose run yields a check that compiles on at least one iteration.
\revision{The prompting baseline draws five independent samples at $T{=}0.8$. Execution-guided repair
uses one run of up to five calls at $T{=}0.2$. Matching call budgets therefore compares two generation
settings, with both sampling and temperature changed. It does not isolate the effect of repair.
Retrieval comparisons within the repair study hold these settings fixed.}

\stitle{Solved@iter and convergence.} solved@iter is the cumulative fraction of instances passing
by iteration $t$ ($t{=}1,\ldots,5$; iteration~1 is the initial generation, 2--5 are repairs),
showing how much of the final pass rate repair contributes; we also report the mean iteration at
which passing checks first pass.

\stitle{Cost per solved check.} We report dollar cost and total token usage (input $+$ output) per
successfully generated check, for each model and configuration. Statistical tests (cluster-robust
GEE and Chi-Square test $\chi^2$) are reported with each result where they are used (Section~\ref{sec:results}).

\section{How Far Repair and Retrieval Reach}
\label{sec:results}

We now evaluate the retrieval-and-repair pipeline, which augments the base description with retrieval and
execution-guided repair (Section~\ref{sec:approach}). 

\subsection{RQ2: How far the retrieval and repair feedback loop reaches}

\revision{With execution-guided repair and retrieval (Table~\ref{tab:lintgen-results}),} the full configuration on \bench compiles
\result{92.0}--\result{99.1\%} of checks and passes \result{56.6\%} (Qwen) to \result{85.0\%}
(Gemini), against a best prompting pass@5 of \result{64.6\%}.
\revision{The Repair-only control estimates retrieval's contribution with the repair budget and
decoding held fixed.}

\revision{\stitle{Repair-only improves on prompting. Retrieval adds smaller gains.}}
Comparing the Repair-only control against the retrieval configurations isolates what
retrieval adds \revision{over Repair-only} (Table~\ref{tab:lintgen-results}).
\revision{Relative to matched-budget prompting, the Repair-only setting is associated with higher
odds of compilation (odds ratio, OR~\result{2.34}) and passing given compilation
(OR~\result{1.82}, Table~\ref{tab:glmm}). These associations include the change in decoding and
sampling.} \revision{Repair-only} already reaches near-ceiling compilation for the
top-performing models (Claude \result{93.8\%}, Gemini \result{95.6\%}) and a high pass rate
\looseness=-1 (\result{66.4\%}, \result{80.5\%}), because the compiler diagnostic lets the model self-correct its
invented API symbols. Adding the API reference then increases pass by only \result{7.1} points for
Claude and \result{1.8} for Gemini, but \result{14.1} for Kimi and \result{6.2} for Qwen, whose
repair-only compilation is lowest (\result{77.0\%}, \result{74.3\%}). Retrieval thus substitutes for
self-correction, large where the model cannot self-correct and small where it can.

\stitle{Modeling approach.} We model success as two conditional stages, the \emph{compile stage}
(whether a check compiles) and the \emph{logic stage} (whether it passes given it compiles), each
fit with a cluster-robust logistic regression (generalized estimating equations, GEE) clustered by
check, since each of the \result{113} checks recurs across the model and configuration
cells~\cite{liang1986,arcuri2011}. Each model contributes a standardized \emph{performance}
covariate, and an API\,$\times$\,performance term tests whether retrieval's value depends on it
(Table~\ref{tab:glmm}).

\begin{table}[t]
\centering
\caption{Two-stage GEE odds ratios on \bench with 95\% CIs. Repair loop vs.\ matched-budget
prompting; retrieval and performance vs.\ Repair-only. \textbf{Bold}: $p<0.05$.}
\label{tab:glmm}
\vspace{-0.5em}
\small\setlength{\tabcolsep}{5pt}
\begin{tabular}{@{}lcc@{}}
\toprule
Predictor & Compile OR & Pass$\mid$compile OR \\
\midrule
\multicolumn{3}{@{}l}{\textit{Repair loop} (vs.\ matched-budget prompting)}\\
\quad Repair loop          & \textbf{2.34}~{\scriptsize[1.71,\,3.21]} & \textbf{1.82}~{\scriptsize[1.37,\,2.42]} \\
\midrule
\multicolumn{3}{@{}l}{\textit{Retrieval and performance} (vs.\ Repair-only)}\\
\quad API reference          & \textbf{2.88}~{\scriptsize[1.93,\,4.30]} & 1.09~{\scriptsize[0.94,\,1.27]} \\
\quad Developer guide        & \textbf{1.52}~{\scriptsize[1.05,\,2.22]} & 1.06~{\scriptsize[0.90,\,1.25]} \\
\quad Performance ($+$1 SD)   & \textbf{2.31}~{\scriptsize[1.83,\,2.91]} & \textbf{1.41}~{\scriptsize[1.23,\,1.63]} \\
\quad API $\times$ performance & 0.76~{\scriptsize[0.55,\,1.06]} & 1.10~{\scriptsize[0.96,\,1.25]} \\
\bottomrule
\end{tabular}
\vspace{-1.5em}
\end{table}

\stitle{Retrieval helps only at the compile stage, dominated by the API reference.}
The API reference more than doubles the odds a check compiles (OR~\result{2.88}, 95\% CI~[1.93,\,4.30],
$p<10^{-4}$), the largest retrieval effect in the study. 
The descriptive gains are largest for the models with the lowest compile rates, but the API $\times$ performance interaction is not significant (OR 0.76, 95\% CI [0.55,\,1.06]). 
The mechanism is concrete: the reference supplies the scanner-callback signatures the description omits, removing the
symbol-hallucination and wrong-override errors that dominate compile failures. The developer guide
has a smaller compile-stage effect
(OR~\result{1.52}, 95\% CI~[1.05,\,2.22]) and none at the logic stage (pass$\mid$compile
OR~\result{1.06}); it is rarely retrieved, clearing the 0.6 similarity threshold for only 9\% of
prompts against 100\% for the API index.

\stitle{Stacking the two sources does not compose.}
Combining both sources does not significantly go beyond the API reference alone; the differences among the retrieval
configurations sit within sampling noise on pass rate, consistent with the null logic-stage effects
in Table~\ref{tab:glmm}.

\stitle{Retrieval helps checks compile.} The API reference's effect
on passing is almost fully mediated by the compile stage: it raises the compile rate but leaves the pass
rate \emph{among checks that compile} unchanged. What governs the conditional pass rate is model
performance (OR~\result{1.41}).

\finding{The repair loop drives most of the gain (compile OR~\result{2.34}, pass OR~\result{1.82}).
Retrieval helps only at compile time, driven by the API reference (OR~\result{2.88}), and its benefit is largest for weaker models.  
The remaining logic gap scales with the model's coding capabilities.}

\begin{figure}[t]
\centering
\includegraphics[width=\columnwidth]{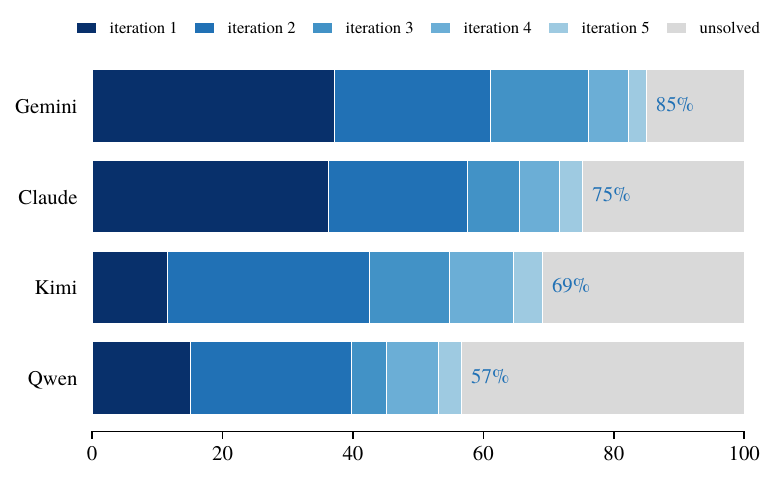}
\vspace{-2em}
\caption{Outcome composition by model, full configuration on \bench (\result{113} checks). Bars
split each model's checks by the iteration that first passed (1 = initial, 2--5 = repairs); the gray
tail is unsolved within the five-call budget.}
\label{fig:budget}
\vspace{-1em}
\end{figure}

\subsection{RQ3: Why do checks fail?}
\revision{\stitle{Adding repair iterations yields diminishing returns.} We fit the cumulative pass
rate across all 452 full-configuration runs (four models $\times$ 113 checks), pooling the
per-model outcomes in Figure~\ref{fig:budget}. It rises from 25.0\% (113/452) on the first call to
50.2\% (227/452) on the second and 71.5\% (323/452) by the fifth. The fifth call adds about four
percentage points.}
The top-performing models solve most checks at the initial generation, while the low-performing
models depend on the repair calls and leave a larger share unsolved (\result{15\%} to \result{43\%}).
\revision{The fitted 82\% asymptote is an extrapolation beyond this observed budget.
In a follow-up, we extended Gemini's 16 compiled-but-failing checks to ten calls. None passed on
calls six through ten. This supports the five-call budget for that subset, but leaves the effect
of longer runs on other models and failures unmeasured.}

\stitle{The API reference clears the compile failures.
\revision{Repair fixes 43.6\% of the compiled-but-failing checks.}} Adding the API reference moves the residual from compilation
to logic. Without it, 54 of 452 runs never compile; the API reference cuts that to 17, while
logic-stage failures stay flat across settings (108--116 compiled checks), consistent with the null
Stage-2 effects in Table~\ref{tab:glmm}. Repair then fixes part of that residual: among the
1{,}261 checks that compile, test feedback increases the pass rate from 52.7\% to 73.4\%, repairing 260
of 596 compiled-but-failing checks (43.6\%). The repair rate rises with performance (Qwen
29.8\% to Gemini 55.0\%; $\chi^2$ $p{=}0.0001$). The remaining gap is therefore semantic, not a
matter of API grounding.

\stitle{The failures are mostly due to under-detection.} After both repair
phases, \result{27\%} of compiled checks still never pass on \bench (\result{441} of
\result{1{,}646}). We observe that (Table~\ref{tab:whyfail}), \result{87\%} are due to \emph{unsoundness}: they compile and run but miss the issue they should report (false negatives), split
between predicates that never flag (\result{48.1\%}) and ones that flag too narrowly
(\result{38.5\%}). The pattern holds across all four models (\result{85}--\result{89\%})
and is the dominant residual for the top models in the best configuration: Gemini's \result{16} remaining
failures are \result{100\%} under-detection and Claude's \result{26} are \result{85\%}, with no
compile or over-detection errors. Reaching near-complete pass thus needs the logic to capture the
targeted issue on checks that already run. Compile failures are also somewhat more recoverable than logic ones (\result{58.4\%} vs.\ \result{53.5\%} eventually repaired):
a compile error names the missing symbol and repair supplies it, whereas a logic failure only signals
that some inputs are mishandled, leaving the model to infer the fix from the rule itself, with no
external artifact to retrieve.

\begin{table}[t]
\centering
\setlength{\tabcolsep}{4pt}
\caption{Failure modes of compiled-but-failing checks on \bench. \emph{All}: pooled over four models
and conditions. \emph{Claude}/\emph{Gemini}: the two models at $\geq$99\% compile in the full
configuration, isolating residual issue-detection logic. Percentages within column; $n$ is the
compiled-but-failing count.}
\vspace{-0.5em}
\label{tab:whyfail}
\begin{tabular}{@{}lrrr@{}}
\toprule
 & & \multicolumn{2}{c}{$\geq$99\% compile} \\
\cmidrule(lr){3-4}
Failure mode & All & Claude & Gemini \\
\midrule
False negative (under-detection)      & \textbf{383 (86.8\%)} & \textbf{22 (84.6\%)} & \textbf{16 (100\%)} \\
\quad\emph{reports nothing}           & 212 (48.1\%) & 11 (42.3\%) & 7 (43.8\%) \\
\quad\emph{catches some, misses some} & 170 (38.5\%) & 11 (42.3\%) & 9 (56.2\%) \\
\quad\emph{flags wrong target}        & 1 (0.2\%)    & 0           & 0 \\
Wrong message / location              & 16 (3.6\%)   & 1 (3.8\%)   & 0 \\
Runtime error                         & 25 (5.7\%)   & 3 (11.5\%)  & 0 \\
Other                                 & 17 (3.9\%)   & 0           & 0 \\
\midrule
Compiled-but-failing ($n$)            & 441          & 26          & 16 \\
\bottomrule
\end{tabular}
\vspace{-1em}
\end{table}

\revision{\stitle{Failure overlap across models.} In the full configuration, the failure analysis
includes 26/16/43/26 compiled-but-failing checks for Claude/Gemini/Qwen/Kimi, respectively.
They cover 62 distinct checks: 34 fail in one model, 14 in two, seven in three, and seven in all
four. Thus, 55/62 (89\%) are not shared compiled-but-failing cases across all models. These counts
describe which checks fail, not whether the models make the same detection error. The seven shared
failures warrant further inspection of both the descriptions and generated logic.}

\begin{table}[H]
\centering
\caption{Pass rate by analysis surface and difficulty tier on \bench, pooled over conditions and
models. ``Multiple'' = $>$1 scanner interface. ORs vs.\ the reference category (cluster-robust);
$^{*}p<0.05$.}
\label{tab:difficulty}
\vspace{-0.5em}
\setlength{\tabcolsep}{6pt}
\begin{tabular}{@{}lcc@{}}
\toprule
Subset & Pass rate & Pass odds vs.\ ref. \\
\midrule
\multicolumn{3}{@{}l}{\emph{By analysis surface (scanner interface)}}\\
\quad XML resource              & 72.2\% & -- \\
\quad Source code (UAST)        & 66.3\% & 0.76\phantom{*} \\
\quad Multiple scanners         & 55.3\% & 0.48* \\
\midrule
\multicolumn{3}{@{}l}{\emph{By benchmark difficulty tier (vs.\ Easy)}}\\
\quad Easy                      & 68.6\% & -- \\
\quad Medium                    & 69.2\% & 1.03\phantom{*} \\
\quad Hard                      & 50.0\% & 0.46* \\
\bottomrule
\end{tabular}
\vspace{-1em}
\end{table}

\begin{figure}[t]
\centering
\begin{tikzpicture}
\begin{axis}[
  ybar, width=\columnwidth, height=4.9cm,
  bar width=5.5pt, enlarge x limits=0.18,
  ymin=0, ymax=100, ylabel={Pass rate (\%)},
  ylabel style={font=\small}, yticklabel style={font=\scriptsize},
  symbolic x coords={Qwen,Kimi,Claude,Gemini},
  xtick=data, xticklabel style={font=\scriptsize},
  ymajorgrids, grid style={gray!18}, axis line style={gray!55},
  tick style={gray!55},
  legend style={font=\scriptsize, at={(0.5,1.03)}, anchor=south,
                legend columns=3, draw=none, fill=none,
                /tikz/every even column/.append style={column sep=7pt}},
  nodes near coords, nodes near coords style={font=\tiny, /pgf/number format/precision=0, /pgf/number format/fixed},
  every node near coord/.append style={yshift=-1pt},
]
\addplot[fill=green!58!black!75, draw=none] coordinates {(Qwen,58)(Kimi,68)(Claude,79)(Gemini,89)};
\addplot[fill=orange!82, draw=none]         coordinates {(Qwen,59)(Kimi,72)(Claude,74)(Gemini,84)};
\addplot[fill=red!72!black, draw=none]       coordinates {(Qwen,43)(Kimi,57)(Claude,71)(Gemini,79)};
\legend{Easy, Medium, Hard}
\end{axis}
\end{tikzpicture}
\vspace{-1em}
\caption{Pass rate by difficulty tier under full configuration.}
\label{fig:difficulty}
\vspace{-1em}
\end{figure}
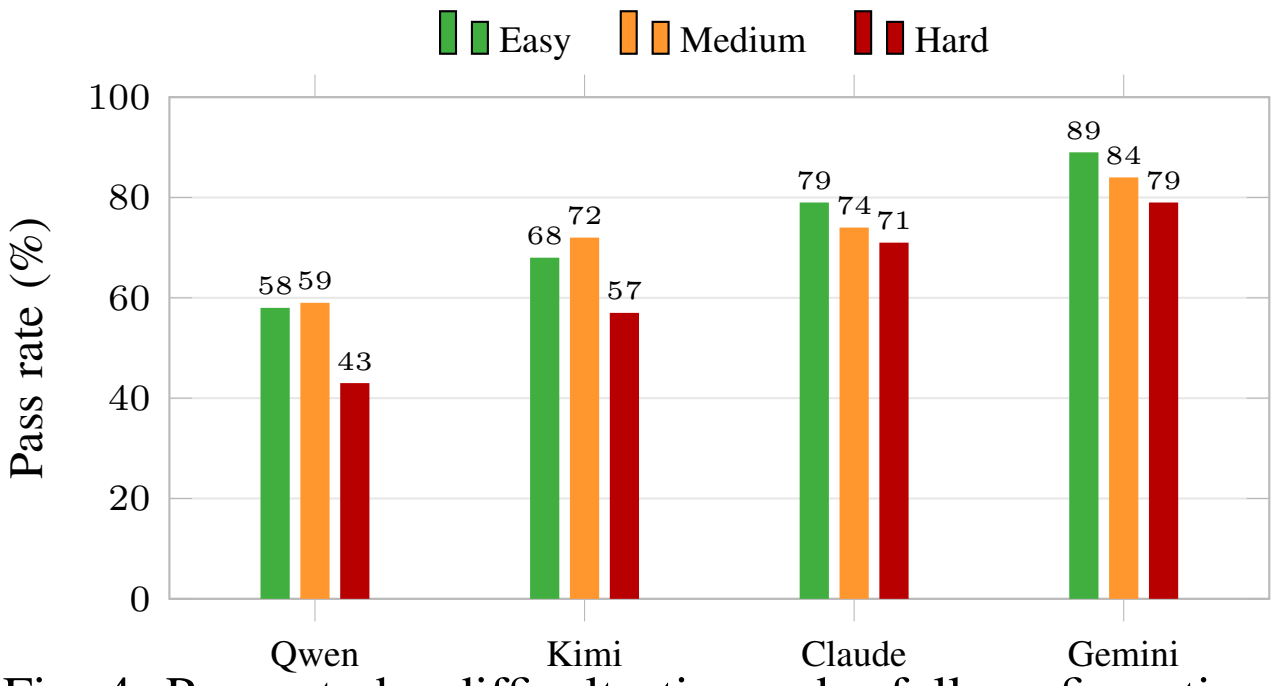

\stitle{What makes a check hard: spanning multiple scanner interfaces.}
\label{sec:difficulty}%
Difficulty concentrates unevenly, and where tells us about the task (Table~\ref{tab:difficulty}).
Counter to the intuition that Java/Kotlin UAST checks are the hardest, source-code checks pass at
\result{66.3\%}, indistinguishable from the XML checks that dominate the benchmark (\result{72.2\%};
OR~\result{0.76}, $p{=}0.34$); difficulty instead concentrates in checks that span \emph{multiple}
scanner interfaces (\result{55.3\%}; OR~\result{0.48}, $p{=}0.017$). Across tiers, pass rate falls
from \textsc{easy} (\result{68.6\%}) to \textsc{hard} (\result{50.0\%}; Figure~\ref{fig:difficulty}),
with only the \textsc{hard} tier significant (OR~\result{0.46}, $p{=}0.014$; \textsc{medium}
indistinguishable from \textsc{easy}, OR~\result{1.03}), so difficulty concentrates in the
\textsc{hard} tier instead of rising evenly. By category, pass rates are lowest on directional or
visual checks (\textsc{rtl} \result{53.1\%}, \textsc{icons} \result{65.6\%}) and highest on
pattern-matching ones (\textsc{security} \result{79.9\%}), though per-category counts are small.

\finding{The generated checks that compile but fail are unsound: \result{87\%} of compiled-but-failing checks are false negatives that miss the issue. 
Retrieval cannot close this gap. It needs issue-detection logic the model must reason out.}

\subsection{RQ4: Cost}
On \bench, cost per solved check spans \$0.12--0.35 and 15--104k tokens (Table~\ref{tab:lintgen-results}). Claude is the cheapest
model in every configuration (\$0.12--0.15, 15--16k tokens), while the top-performing Gemini costs
more (\$0.21--0.23, 27--28k) and Kimi's verbose generations make it the most token-heavy
(56--104k). Turning to configurations, retrieval does not raise cost per solved check despite its
longer context, because a higher solve count offsets the extra input. For Claude, the full
configuration is even the cheapest per success (\$0.120, 14.6k). Holding the model fixed, the API
reference is at least as cost-effective as the developer guide, and cheaper for the lowest-performing
model (Qwen \$0.27 vs.\ \$0.35, 50k vs.\ 64k tokens), because guide-only generation needs more repair
to pass. Even Repair-only is no cheaper, since its compiler feedback can itself expand the context.

\finding{Cost per solved check is driven by token count. Claude uses the fewest tokens (15k) and is the cheapest at \$0.12 per solved check, yet it is not the top performer. 
Gemini performs best but costs more.
Retrieval is roughly cost-neutral. Its expanded context length is offset by a higher solve count.}

\section{Discussion}
\label{sec:discussion}

\subsection{What drives generation, and what knowledge adds}
Most of the improvement over prompting appears with execution-guided repair, and retrieval adds less (RQ2). Hallucinated API calls are the dominant baseline failure, and a top-performing model self-corrects them from the
compiler's diagnostics, so the loop alone reaches near-ceiling compilation with no retrieved
knowledge. Retrieval helps in proportion to need: the API reference substitutes for self-correction
at the compile stage, much for weak models and little for strong ones, while the developer
guide adds little and the two do not compose. What matters is execution feedback plus, where needed,
retrieved API grounding, not static scaffolding or prose documentation~\cite{apidocrag}.

\subsection{Where LLMs still fail}
Once a check compiles, the residual is issue-detection logic: the check runs but its predicate misses
the issue (RQ3). For the top models, which all but saturate compilation, it is nearly the entire
remaining gap. Repair fixes some of these failures from the test diagnostics, but the rest need
analysis reasoning that neither retrieved source supplies. The bottleneck thus shifts from knowing
the API to reasoning about the analysis, so we expect further gains from tooling aimed there, such as
program-analysis-aware decoding or counterexample-guided repair.

\subsection{An ablation on description quality}
Human validation doubles as an ablation on description quality. The \result{35} checks it removes are
effectively unsolvable, passing at only \result{12.1\%} under the full configuration, and \result{19}
of the \result{35} are never solved by any model in any configuration. Adding them back, the
\bench-unverified \result{148}-check set, drops the full-configuration pass rate from \result{71.5\%}
on \bench to \result{57.4\%}, a \result{14}-point gap that comes entirely from these low-yield checks. \bench
thus measures capability without these benchmark artifacts.
\revision{Exclusions include under-specification and sibling-test dependencies; low pass rates
do not rule out solutions from other models or prompts.}

\subsection{Implications}
The description-only setting is a realistic entry point for check authoring: from a one-sentence
rule, with no tests written first, the pipeline produces a working, test-passing check for up to
\result{85.0\%} of \bench checks at \$0.12 to \$0.35 each (RQ4), so a developer can describe a rule
and review a working candidate instead of implementing it from scratch. For framework designers, the failure modes point to the API and
documentation gaps that most mislead generation; for researchers, \bench is an objective benchmark
scored by a held-out test oracle.
\revision{Repair used the held-out test oracle; reduced authoring or review effort was not
measured against a human baseline.}

\section{Threats to Validity}
\label{sec:threats}

\stitle{Construct validity.} We equate a correct check with passing the developers' original test
suite, which encodes the intended behavior; a check could still misbehave on unseen code, so the
absence of false positives we report is relative to these tests, and broader validation is future
work. \revision{Relaxed assertions omit message and quick-fix requirements; preserved
\texttt{expectClean()} cases cover only some clean inputs. Empty-detector mutation filtering
rejects a trivial implementation without establishing test completeness (Section~\ref{sec:harness}).}
To confirm that passing reflects real issue-detection logic and not gaming of the held-out
oracle, we compared the \result{1{,}355} detectors generated on \bench against the original implementation. Passing detectors match the original's analysis
scope ($p\!<\!10^{-4}$) and use more of its API vocabulary ($p\!<\!0.01$) than \emph{failing} ones
(Mann--Whitney). We find no passing detector that hardcodes test outputs, so \revision{this form of}
oracle-gaming is effectively absent. \revision{Other overfitting to the fixed tests through
repeated diagnostic feedback remains possible.}

\stitle{Internal validity.} Retrieval and the repair loop could be confounded, since both sit on
top of the prompting baselines; the Repair-only condition separates them
(Table~\ref{tab:lintgen-results}), and our marginal comparisons (API reference vs.\ developer
guide, the performance interaction) are within-design. \revision{In the execution-guided repair study,}
each model runs each check once per configuration (a single execution-guided repair run, not $k$ samples), so we base claims on aggregate rates and the
GEE model, not individual check flips, with fixed throughput~decoding.
\revision{Prompting uses five independent samples at $T{=}0.8$; repair uses up to five calls
at $T{=}0.2$, so matching call budgets leaves sampling and temperature differences. Screening
used one author; annotator agreement and sensitivity to few-shot selection were not measured.}

\stitle{External validity.} We study a single framework (Android Lint) and a finite model set, so
findings may not transfer to other analyzers or models. 
We mitigate by spanning Lint's full range
of analysis representations and four model families.

\stitle{Data contamination.} The AOSP detectors and their tests are public, so they may have been included in the models' pretraining data, and memorization could inflate compile/pass rates. 
Two observations bound this concern. First, if a model were reproducing \revision{an exact, complete copy of}
a memorized detector \revision{matching our evaluated reference}, it would compile and pass perfectly under zero-shot prompting. Instead, checks fail by hallucinating APIs at compile time and by under-detecting at the logic stage.
\revision{Partial memorization can still contribute to these imperfect outputs~\cite{riddell2024contamination}.}
Second, our component comparisons hold the model fixed, so memorization that does not depend on prompt context would shift all conditions equally and leave the differences between Repair-only and the retrieval configurations unchanged. Context can affect memorized output~\cite{carlini2023memorization}, however, so contamination may still contribute differently across the prompting, repair, and retrieval conditions.

\revision{At our May 22, 2026 checkout, the 81 reference files' last-modified dates span
March 9, 2018--March 11, 2026 (median: January 29, 2026). Of these, 41/81 (50.6\%) postdate
Claude's January 2026 cutoff, and 49/81 (60.5\%) Gemini's January 2025 cutoff (Qwen and Kimi do not
disclose cutoff dates). All files were created before both cutoffs, so updates do not establish unseen code.}

\revision{Of 451 available candidates from 452 full-configuration runs (one Qwen run produced none),
none matched the reference after comment/whitespace normalization or reached 0.95 token-sequence
similarity (\texttt{SequenceMatcher}); model means were 0.30--0.32. This weighs against verbatim
reproduction; partial memorization remains possible. Evaluation on new, non-public checks remains future work.}

\stitle{Reliability.} We release \bench, the descriptions, the harness, and
all prompts and generated outputs so that our results can be reproduced and future
check-generation methods scored against the same objective oracle.

\section{Related Work}
\label{sec:related}

\subsection{Traditional static analysis and custom checks}
Hand-written analyzers such as FindBugs~\cite{findbugs}, PMD~\cite{pmd}, and Infer~\cite{infer}
rely on manually authored bug patterns~\cite{ayewah2008using}, and frameworks such as
Soot~\cite{soot} and FlowDroid~\cite{flowdroid} provide the program-analysis machinery for Android
taint tracking; such analysis has been deployed at industrial scale~\cite{distefano2019scaling}.
Authoring custom rules is difficult even for experienced
developers~\cite{mendonca2022custom,christakis2016developers}, and developers frequently avoid
these tools because of API complexity and friction~\cite{johnson2013static,tymchuk2018jit}, which
motivates automated check generation. AutoComply~\cite{autocomply} hand-constructs
checkers for Android Auto compliance that Android Lint does not cover, indicating how often Lint
lags emerging platform features. \bench evaluates whether LLMs can generate such detectors from a natural-language description alone.

\subsection{LLMs for code and static analysis}
Pre-trained code models such as
CodeBERT~\cite{codebert}, CodeT5+~\cite{codet5p}, StarCoder~\cite{starcoder}, Code
Llama~\cite{codellama}, and DeepSeek-Coder~\cite{deepseekcoder} underpin their use across software
engineering~\cite{hou2024large,fan2023large}, but most work applies LLMs \emph{to} analysis, not
to building analyzers: interleaving analyzer calls with LLM queries where the analyzer cannot
proceed~\cite{interleave}, integrating LLMs into bug-detection
pipelines~\cite{li2024enhancing}, inferring source--sink specifications for
vulnerability detection~\cite{iris}, auditing privacy compliance over a curated Android-API
knowledge base~\cite{priagent}, and benchmarking LLMs against traditional tools for vulnerability
detection~\cite{llmvsstatic,vulndet,zhou2024llmvuln}, code-smell detection~\cite{codesmell}, and
specification mining~\cite{specmining}. Related studies ask whether code LLMs perform
static-analysis reasoning~\cite{docodellms} and contextualize them with program-analysis
insights~\cite{cldk}. These works use LLMs as analyzers; we instead study LLMs as builders of the
analyzer itself.

\subsection{LLM-based synthesis of static-analysis artifacts}
A few systems synthesize analysis artifacts with LLMs. KNighter
generates Clang Static Analyzer checkers from historical bug-fix patches~\cite{knighter}, and
QLCoder synthesizes CodeQL queries from CVE metadata with retrieval and execution
feedback~\cite{qlcoder}. Closest to us, AutoChecker generates PMD checkers from a rule description
\emph{and a test suite}, repairing against the supplied tests~\cite{autochecker}. \revision{Our task specification is a description, with test source hidden from the model. Lint adds a representation-selection problem: Clang checkers inspect C/C++ syntax trees and CodeQL queries operate on a source-derived relational database, while Lint exposes scanners for source, resources, manifests, and build scripts. Lint authors also need Android framework classes, resource conventions, manifest schemas, and Gradle DSL knowledge~\cite{lintApiGuide,lintUserGuide}.} We also contribute \bench, which scores check
\emph{generation} by compiling each candidate and running the check's own held-out tests. It thus
differs from benchmarks that evaluate analysis or repair of existing code, such as the
vulnerability-detection benchmark CWE-Bench-Java~\cite{iris} and the mobile issue-resolution
benchmark MobileDev-Bench~\cite{mobiledevbench}, and from work that automates linter
\emph{configuration} of existing rules~\cite{lintcfg}, not synthesis of new ones.

\subsection{Execution-guided self-repair and retrieval}
\bench combines two techniques. The first is execution-guided self-repair, in which a model revises
its code from compiler or test feedback~\cite{chen2024selfdebugging,shinn2023reflexion}, although
self-repair does not always improve results~\cite{olausson2023selfrepair}. The second is
retrieval: retrieval-augmented generation grounds models in knowledge beyond their
parameters~\cite{lewis2020rag}, and for code it can supply the specific APIs a task requires,
including through compositional API recommendation~\cite{ma2024compositional}. For unfamiliar APIs,
retrieving relevant documentation, especially usage examples, can substantially raise pass
rates~\cite{apidocrag}, though the benefit is not guaranteed: in checker synthesis, KNighter
reports that retrieval over checker examples matched fixed few-shot examples while roughly doubling
token cost, and so retained the fixed examples~\cite{knighter}. Whether, and at what cost,
retrieval helps check generation thus remains open. \bench applies both in a fixed
generate-validate-repair loop specialized for detector synthesis, repairing from harness
diagnostics and retrieving over Lint's API reference and developer guide.

\section{Conclusion}
\label{sec:conclusion}

We studied whether LLMs can generate Android Lint checks from natural language descriptions alone and what external knowledge they need to do so. We introduced \bench, a human-validated benchmark
of \result{113} real-world checks with their original test suites and an automated compile-and-test
harness. We evaluated four LLMs across prompting, retrieval of the Lint API reference and developer
guide, and execution-guided repair. Prompting alone reaches \result{64.6\%} (pass@5); adding
retrieval and execution-guided repair raises this to \result{85.0\%} at a matched five-call budget.
Our results show that execution feedback raises the pass rate, while retrieved API knowledge raises
the compile rate, which reaches up to \result{99\%} and helps the low-performing models most. The
\revision{residual failures are mostly false negatives.} Of the generated checks that compile but fail their tests,
\result{87\%} miss the issue they should report, so getting the issue-detection logic right remains
an open challenge. We release \bench, the harness, and prompts~\cite{artifact} to support future work.

\section*{Acknowledgment}
This material is based upon the work supported in part by the U.S. National Science Foundation (NSF) under Grant numbers~2223826, 2616791, 2616792, and 2616793. Any opinions, findings, and conclusions or recommendations expressed in this material are those of the authors and do not necessarily reflect the views of the NSF.

\balance
\bibliographystyle{IEEEtran}
\bibliography{references}

\end{document}